\documentclass[10pt,a4paper]{scrartcl}
\usepackage{amsmath,amssymb}
\usepackage{authblk}

\usepackage[T1]{fontenc}
\usepackage{lmodern}
\usepackage{geometry}
\usepackage{setspace}
\usepackage{graphicx}
\usepackage[dvipsnames,svgnames,x11names]{xcolor}
\usepackage{booktabs}
\usepackage{fancyvrb}
\usepackage[all]{nowidow}        
\usepackage[colorlinks=true,linkcolor=black!70, citecolor=BrickRed, urlcolor=Navy]{hyperref}

\usepackage{tikz}
\usetikzlibrary{positioning,shadows}

\usepackage[style=numeric, backend=biber, giveninits=true]{biblatex}
\AtEveryBibitem{%
  \iffieldundef{doi}{}{\clearfield{url}}%
}

\addtokomafont{section}{\sffamily\large}
\addtokomafont{subsection}{\sffamily\normalsize}

\usepackage[most]{tcolorbox}
\usepackage{fontawesome5}
\colorlet{humancolor}{Indigo!30}      
\colorlet{agentcolor}{Teal!30}        
\colorlet{toolcolor}{LightCoral!20!gray!15}    
\colorlet{analysiscolor}{Goldenrod!30}

\colorlet{humancolorlight}{humancolor!40!gray!90}
\colorlet{agentcolorlight}{agentcolor!40!gray!90}
\colorlet{toolcolorlight}{toolcolor!40!gray!90}
\colorlet{analysiscolorlight}{analysiscolor!40!gray!90}

\tcbset{
  dialogue/.style={
    enhanced,
    colback=white!98!gray,        
    colframe=black!15,    
    boxrule=0.6pt,
    arc=2pt,
    left=6pt,
    right=6pt,
    top=6pt,
    bottom=6pt,
    titlerule=0pt,
    coltitle=black!80,
    fonttitle=\bfseries\small,
    fontupper=\sffamily\footnotesize\color{black!80},
    fontlower=\sffamily\footnotesize,
  }
}
\newtcolorbox{humanbox}[1][]{
  dialogue,
  sharp corners=southeast,arc=10pt,
  drop shadow={humancolorlight},
  colbacktitle=humancolor,   
  title=\faUser~Human (query),
  #1
}

\newtcolorbox{agentbox}[1][]{
  dialogue,
  sharp corners=southwest,arc=10pt,
  drop shadow={agentcolorlight},
  colbacktitle=agentcolor,  
  title=\faRobot~JutulGPT (response),
  #1
}

\newtcolorbox{toolbox}[1][]{
  dialogue,
  sharp corners=southwest,arc=10pt,
  drop shadow={toolcolorlight},
  colbacktitle=toolcolor, 
  title=\faTools~Tool calls,
  listing only,
  listing options={
    basicstyle=\ttfamily\footnotesize,
    breaklines=true,
  },
  #1
}

\newtcolorbox{analysisbox}[1][]{
  dialogue,
  sharp corners=southwest, arc=10pt,   
  drop shadow={analysiscolorlight},
  colbacktitle=analysiscolor,
  title=\faBrain~Agent interpretation,             
  #1
}

\begin{document}

\title{Agentic Scientific Simulation: Execution-Grounded Model Construction and Reconstruction}
\author[1]{Knut--Andreas Lie}
\author[1]{Olav Møyner}
\author[2]{Elling Svee}
\author[1]{Jakob Torben}
\affil[1]{\large SINTEF Digital, Oslo, Norway}
\affil[2]{\large Department of Mathematical Sciences, NTNU, Trondheim, Norway}
\date{}

\maketitle

\begin{abstract}
We ask whether an LLM agent can serve as a genuine scientific assistant for physics-based simulation: given a natural-language description of a model, can it produce an executable, physically valid simulation configuration that faithfully represents the user's intent, navigating API contracts, unit conventions, and global constraints in the way a knowledgeable research assistant would? The challenge is not merely generating code that runs or even satisfies physical consistency tests. Natural-language descriptions of simulation models are inherently underspecified, and different admissible resolutions of implicit choices produce physically valid but scientifically distinct configurations. Without explicit detection and resolution of these ambiguities, neither the correctness of the result nor its reproducibility from the original description can be assured.

This paper investigates \emph{agentic scientific simulation}, where model construction is organized as an execution-grounded interpret--act--validate loop, and the simulator serves as the authoritative arbiter of physical validity rather than merely a runtime. We present \texttt{JutulGPT}, a reference implementation built on the fully differentiable Julia-based reservoir simulator \texttt{JutulDarcy}. The agent combines structured retrieval of documentation and examples with code synthesis, static analysis, code execution, and systematic interpretation of solver diagnostics. Underspecified modelling choices are detected explicitly and resolved either autonomously (with logged assumptions) or through targeted user queries.

The results demonstrate that agent-mediated model construction can be grounded in simulator validation while also revealing a structural limitation: choices resolved tacitly through simulator defaults are invisible to the assumption log and therefore to any downstream representation. A secondary experiment, autonomous reconstruction of a reference model from progressively abstract textual descriptions, shows that reconstruction variability exposes latent degrees of freedom in simulation descriptions and provides a practical methodology for auditing reproducibility. All code, prompts, and agent logs are publicly available.
\end{abstract}


\section{Introduction: Execution-Grounded Agents in Scientific Simulation}

Recent progress in large language models (LLMs) and agentic orchestration has enabled tool-using code assistants and long-horizon coding agents that can iteratively generate, execute, and repair nontrivial software artifacts. These systems can inspect entire repositories, modify files, run commands, and refine implementations through repeated interaction, substantially reducing the effort required to prototype and explore simulation scripts.

Two paradigms have emerged for LLM-assisted software development~\cite{Sapkota2025}. \emph{Vibe coding} emphasizes intuitive, human-in-the-loop interaction: the user states intent conversationally and the model generates code, with successful execution as the primary correctness criterion. \emph{Agentic coding} goes further, enabling autonomous multi-step planning, tool use, and iterative self-repair with minimal human intervention per cycle. Capable agentic systems can already construct meaningful correctness tests autonomously, such as conservation checks and patch tests for physics-based components, and iterate until these are satisfied. For scientific simulation, however, a deeper challenge remains: natural-language descriptions of simulation models are inherently underspecified, and different admissible resolutions of implicit choices produce physically valid but scientifically distinct configurations. The question of whether a completed simulation faithfully represents the user's intent, and whether it can be reproduced from its description, is not resolved by execution success alone.

Translating a conceptual description of a physical scenario into an executable simulation model is thus not merely a programming exercise. It involves selecting governing equations, specifying constitutive relations, defining boundary and initial conditions, choosing discretizations and solver settings, and resolving underspecified modeling assumptions. Even within modern modular simulator frameworks, this process typically proceeds through iterative refinement, inspection of solver diagnostics, and clarification of implicit choices.

These considerations motivate the emerging direction of \emph{agentic scientific simulation}. The key insight is that a physics-based simulator, by virtue of its internal enforcement of governing equations, constitutive consistency, and solver convergence, is already equipped to act as an authoritative arbiter of physical validity. Embedding an LLM within a structured interpret--act--validate loop around such a simulator allows this enforcement capability to be exploited directly: the agent mediates between high-level modelling intent and executable specifications, while the simulator determines what is physically admissible.\looseness=-1

In this work, we explore this direction through a concrete reference implementation, \texttt{JutulGPT}, built on top of the fully differentiable Julia-based simulator \texttt{JutulDarcy} \cite{Møyner2025}. The objective is not to demonstrate LLMs' capability of writing simulation code, but to investigate how an agent can mediate scientific model construction under underspecification, solver feedback, and reproducibility requirements.

To assess this perspective systematically, we evaluate the framework along four dimensions. First, we examine semantic navigation of simulator documentation: whether the agent can retrieve, organize, and synthesize distributed documentation into coherent representations of available functionality. Second, we study execution-grounded model construction and repair: whether the agent can converge to executable, physically consistent models through iterative interaction with simulator execution and solver diagnostics. Third, we analyze structured ambiguity detection and resolution, investigating how underspecified modeling choices are surfaced, logged, and resolved through the interpret--act--validate loop. Fourth, we study reconstruction fidelity under representational compression, where executable models are regenerated from progressively abstract textual descriptions. Together, these probes are intended to clarify both the capabilities and the limitations of execution-grounded agentic workflows in scientific simulation.\looseness=-1

Recent conceptual work has emphasized artificial intelligence (AI) systems as \emph{scientific collaborators} rather than passive tools, integrating reasoning, execution, and interpretation in closed loops that mirror aspects of scientific practice \cite{Wei2025}, ranging from autonomous experiment execution and validation \cite{Park2026} to end-to-end scientific discovery including manuscript authorship \cite{Yamada2025}. In parallel, benchmark efforts such as \emph{SciCode} \cite{Tian2024} have highlighted both the potential and the limitations of LLMs in generating scientific code, underscoring the importance of correctness, interpretability, and domain grounding.

A growing number of domain-specific systems couple large language models directly to established solver ecosystems. For multiphysics problems based on MOOSE, MooseAgent \cite{Zhang2025} combines an LLM with structured retrieval of annotated input files and iterative verification against solver feedback. In computational mechanics and finite-element modeling, systems such as AutoFEA \cite{Hou2025}, MechAgents \cite{Ni2024}, and FeaGPT \cite{FeaGPT2025} demonstrate that conversational or multi-agent workflows can coordinate meshing, boundary-condition specification, solver execution, and error correction. Similar developments have emerged in computational fluid dynamics, where OpenFOAM-based agents integrate retrieval-augmented generation with solver execution to support interactive case construction and modification \cite{Pandey2025,Chen2024,Chen2025,Chen2026,Xu2024}, with recent systems additionally accepting multimodal inputs that parse geometry directly from images \cite{Yang2025}. Dataset-oriented efforts such as NL2FOAM \cite{ZDong2025} and interactive systems such as ChatCFD \cite{Fan2025} further illustrate the feasibility of mapping natural-language descriptions to executable solver configurations. Li et al.~\cite{Li2025} target PDE solver generation more broadly, using LLMs to produce numerical method code directly from problem statements.

These systems demonstrate that LLMs can be embedded in end-to-end simulation workflows when tightly coupled to well-documented simulator architectures. However, much of the existing work emphasizes translation from text to solver input or iterative correction of syntactic and configuration errors. The present work differs in focusing explicitly on execution-grounded scientific model construction, where ambiguity detection, assumption logging, and simulator-enforced physical validation are treated as key components of the workflow. Rather than centering on the generation of valid input files, we investigate how the interpret--act--validate loop can mediate underspecified modeling tasks and how executable simulator state relates to its textual representations.


\section{Simulation Workflows and Agentic Interfaces}
\label{sec:background}

Modern scientific simulators typically expose modular abstractions for grids, discretizations, physical properties, and linear and nonlinear solvers, enabling complex models to be assembled from reusable components. Nevertheless, translating a conceptual modeling objective into a consistent and executable solver configuration remains a nontrivial task. An agentic interface supports this process by mediating between high-level modeling intent and simulator specifications through an execution-grounded loop of construction, execution, and revision. Success still depends on aligning governing equations, constitutive closures, boundary and initial conditions, and solver controls in a manner that is both physically admissible and numerically stable.\looseness=-1


Many multiphysics applications involve coupled physical mechanisms operating across scales. Examples relevant to the present work include subsurface energy systems, which couple flow and heat transport, and electrochemical models, which couple ionic transport, charge conservation, and thermal effects. These examples reflect domains represented within the \texttt{Jutul} ecosystem and the authors' research context, but the underlying challenges are common to PDE-based simulation frameworks more broadly. In such settings, model construction involves selecting and assembling compatible balance laws and closure relations while respecting solver-specific constraints. An agentic interface operating in this environment must therefore reason not only about syntax but also about dependencies among physical components and admissible combinations of constitutive models.


Embedding an agent within a scientific simulation workflow introduces additional requirements. Generated configurations must be traceable and versioned to ensure reproducibility. Numerical stability and convergence behavior must be respected, and solver diagnostics must be interpreted in a structured manner. Constitutive assumptions and default closures must be surfaced explicitly to avoid silent changes in the governing model. These constraints motivate architectures in which the simulator enforces physical and numerical admissibility while the agent mediates interpretation and refinement.


\section{The \texttt{Jutul} Ecosystem}
\label{sec:Jutul}

\texttt{Jutul}\footnote{Source code available at \href{https://github.com/sintefmath/Jutul.jl}{github.com/sintefmath/Jutul.jl}} is an open, modular framework for composable and fully differentiable simulation of systems governed by partial differential equations (PDEs). It provides the computational substrate on which \texttt{JutulGPT} operates. The framework is designed around explicit abstractions for balance laws, constitutive closures, discretization operators, and nonlinear solvers, enabling models to be assembled from interoperable components while preserving transparency of the resulting executable state. 

\texttt{Jutul} is designed from the ground up around equations expressed in residual form, which allows automatic differentiation to permeate the entire computational stack. This serves two purposes. First, it automates linearisation and Jacobian assembly for the Newton solver, eliminating the need for hand-coded derivatives of physical operators. Second, and more distinctively, it enables computation of sensitivities with respect to essentially any parameter in the system---not only physical model parameters such as constitutive closure coefficients, but also parameters that govern the construction of the numerical model itself, such as grid spacing and timestep sizes. The latter capability goes substantially beyond what is available in most simulators, where discretisation parameters are treated as fixed infrastructure outside the reach of the differentiation machinery. 

\subsection{Balance--closure formulation}

Across all branches of the ecosystem, models are expressed as coupled systems of balance equations and closure relations. In continuous form,
\begin{align}
  \frac{\partial A(U)}{\partial t}
  + \nabla \cdot F(U, W, x, t)
  &= Q(U, W, x, t),
  \label{eq:jutul-balance}
  \\
  C(U, W, x, t; \theta) &= 0,
  \label{eq:jutul-closure}
\end{align}
where $U$ denotes primary state variables, $W$ secondary variables and parameters, $A$ accumulation terms, $F$ fluxes, $Q$ sources, and $C$ constitutive or algebraic closure relations parameterized by $\theta$. This decomposition provides a common structural template for diverse physical models while keeping balance laws and constitutive assumptions explicitly separated.

In practice, these equations are discretized using finite-volume methods in space and implicit time integration. The resulting nonlinear residual equations are assembled from modular balance and closure contributions and solved using Newton-type methods with sparse linear solvers. Because each physical mechanism contributes explicitly to the residual and Jacobian, new processes or constitutive models can be introduced without restructuring the surrounding solver infrastructure.

\subsection{Domain branches in \texttt{Jutul}}

The same abstractions support multiple domain-specific simulators. \texttt{JutulDarcy} implements single-phase, multiphase, and compositional flow in porous media. \texttt{Fimbul} extends this framework to geothermal applications that include energy conservation \parencite{Klemetsdal2025}. \texttt{BattMo} targets electrochemical systems, coupling ionic transport and charge conservation \parencite{Clark2026}. \texttt{VOCSim} and \texttt{Mocca} address gas emissions and adsorption processes, respectively. Although these branches target different application domains, they share common data structures, discretization operators, solver strategies, and differentiation mechanisms.

\subsection{Properties relevant to agentic mediation}

Several characteristics of \texttt{Jutul} are particularly relevant for execution-grounded agentic interaction. The framework exposes explicit type structures, well-defined constructor interfaces, and structured error messages. All components are open source and documented, allowing documentation, docstrings, and example scripts to be indexed and retrieved systematically. Because balance laws and closures are represented explicitly in code, constitutive assumptions and parameterizations are inspectable rather than implicit in opaque input formats.

In addition, the residual-based architecture and structured diagnostics enable precise interpretation of runtime failures. Constructor mismatches, inconsistent parameterizations, and convergence issues are surfaced through explicit error objects and solver logs, which can be parsed and fed back into the interpret--act--validate loop.

Together, these properties make \texttt{Jutul} a suitable substrate for studying execution-grounded agentic workflows: the simulator enforces physical and numerical admissibility, while its modular and introspectable design allows the agent to retrieve, construct, execute, and revise models in a controlled and reproducible manner.


\section{The \texttt{JutulGPT} Framework}
\label{sec:JutulGPT}

To investigate execution-grounded agentic model construction in a concrete setting, we implemented \texttt{JutulGPT}, a reference implementation that couples an LLM to \texttt{JutulDarcy}. Full source code, prompts, and example logs are publicly available at \href{https://github.com/SINTEF-agentlab/JutulGPT}{github.com/SINTEF-agentlab/JutulGPT}. The framework operationalizes the interpret--act--validate loop and provides a controlled environment for studying ambiguity resolution, solver-mediated refinement, and reconstruction under underspecification.

\subsection{Execution-grounded workflow}
At a high level, \texttt{JutulGPT} organizes model construction as an iterative loop consisting of three phases:
\begin{itemize}
    \item \textbf{Interpret:} Parse user intent, identify underspecified modeling choices, and determine required simulator components.
    \item \textbf{Act:} Retrieve relevant documentation and examples, generate or modify Julia code, and perform static checks.
    \item \textbf{Validate:} Execute the generated code in a local Julia environment, analyze solver diagnostics and runtime errors, and determine whether physical and numerical admissibility criteria are satisfied.
\end{itemize}

If validation fails, the agent revises the implementation and re-enters the loop. Ambiguities detected during interpretation are either resolved autonomously (with explicit logging of assumptions) or escalated to the user through targeted clarification queries. Termination occurs when the simulator runs to completion and no outstanding ambiguity remains. Because JutulDarcy enforces conservation tolerances, closure consistency, and solver convergence internally, a completed run is a simulator-grounded validation: the simulation would not have run to completion had these conditions not been met. This is a strictly stronger criterion than the absence of a runtime error.

This structure distinguishes the workflow from one-shot text-to-code generation and from generic coding assistants that optimize primarily for syntactic correctness. The simulator acts as an authoritative truth engine: execution feedback, convergence diagnostics, and error messages directly constrain the space of admissible implementations.

\subsection{System architecture}
The \texttt{JutulGPT} framework comprises three interacting layers:
\begin{itemize}
    \item The \emph{conversational layer}, which manages dialogue state, intent parsing, and ambiguity detection.
    \item The \emph{tool layer}, which exposes structured capabilities for documentation retrieval, code analysis, file manipulation, and execution.
    \item The \emph{runtime layer}, which interfaces directly with \texttt{JutulDarcy}, executes generated scripts, and returns structured diagnostics.
\end{itemize}
Communication between these layers implements the loop described above. Retrieval operations are grounded in indexed simulator documentation, docstrings, and curated example scripts. Generated code is first subjected to static analysis before execution in a local Julia environment. Runtime errors and solver diagnostics are captured, parsed, and returned to the agent, enabling targeted revision rather than repeated trial-and-error re-generation.
\begin{figure}
    \centering
    \includegraphics[width=0.8\linewidth]{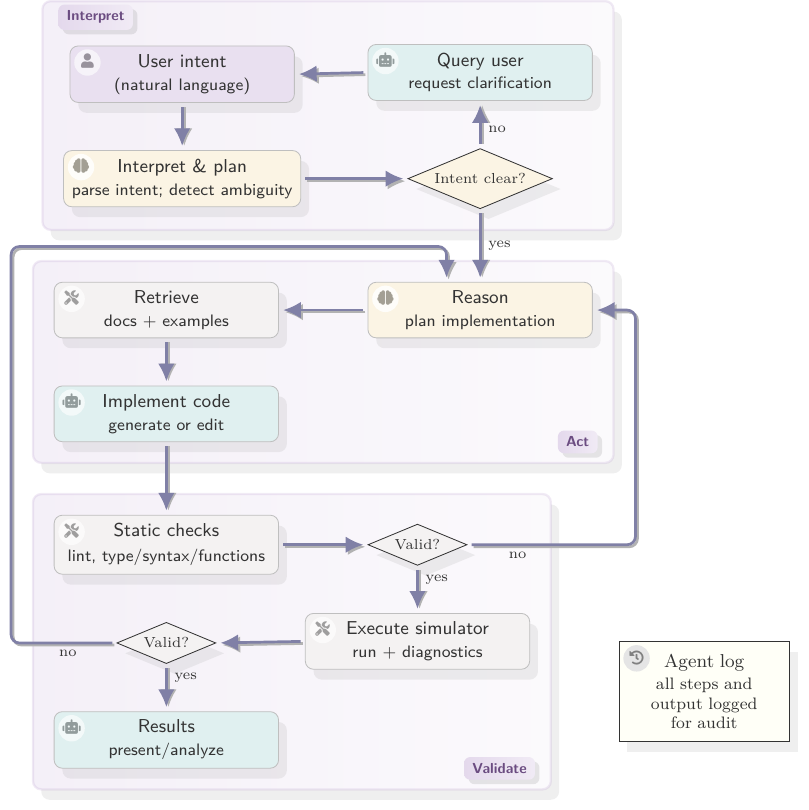}
    \caption{Typical iterative interpret--act--validate loop used by \texttt{JutulGPT}. The agent incrementally interprets user intent, detects ambiguities, and resolves them either autonomously (with explicit assumption logging) or via targeted user queries. Code generation is grounded in retrieved documentation and validated through static analysis, execution and simulator diagnostics, with failures triggering revision cycles. The loop terminates when the simulator runs to completion; an event that constitutes a validity certificate by virtue of JutulDarcy's internal enforcement of conservation tolerances, closure consistency, and solver convergence.}
    \label{fig:jutulgpt-loop}
\end{figure}
All prompts, tool invocations, generated code fragments, and execution outputs are logged. This logging discipline supports traceability of assumptions and reproducibility of the modeling workflow.

Figure~\ref{fig:jutulgpt-loop} illustrates the operational behavior of the framework as an iterative cycle. The interpret phase extracts modeling intent and identifies missing specifications. The act phase performs retrieval and code synthesis grounded in simulator documentation and examples. The validate phase executes the generated configuration and evaluates solver diagnostics and runtime errors. Execution feedback directly constrains subsequent revisions, and the loop continues until a physically and numerically admissible configuration is obtained.

In this mapping, the interpret phase primarily engages the conversational layer, where intent is parsed and missing specifications are identified. The act phase spans the conversational and tool layers, combining retrieval with code synthesis and modification. The validate phase interfaces with the runtime layer, where generated configurations are executed and assessed using structured solver diagnostics. Validation outcomes may trigger implementation-level revisions or renewed clarification of modeling assumptions. This three-phase abstraction thus enforces a clean separation of responsibilities across layers, with each phase having a well-defined scope and a defined interface to the next,

\begin{table}
    \centering
    \caption{Overview of tools available to JutulGPT.}
    \label{tab:JutulGPTToolOverview}\small
    \begin{tabular}{ll}
    \toprule
    \multicolumn{2}{l}{\textbf{Documentation Retrieval}} \\
    \midrule
    Keyword search & Retrieves documentation using keyword matching. \\
    Docstring lookup & Displays function, class, or module docstrings. \\
    Example search & Finds examples related to a query. \\
    \midrule
    \multicolumn{2}{l}{\textbf{Code Analysis}} \\
    \midrule
    Linter & Analyzes code for errors, style issues, and common pitfalls. \\
    Code runner & Executes code snippets in a controlled environment. \\
    Terminal commands & Runs shell commands and returns their output. \\
    \midrule
    \multicolumn{2}{l}{\textbf{Input and Output}} \\
    \midrule
    Read file & Reads the contents of a file. \\
    Write file & Writes data to a file. \\
    List files & Shows the directory structure and available files. \\
    \bottomrule
    \end{tabular}
\end{table}

The agent grounds its reasoning through retrieval over simulator documentation, docstrings, and example scripts, using two complementary retrieval modes. Semantic retrieval-augmented generation (RAG) helps the agent map user intent to relevant working configurations, even when the user’s wording does not match simulator terminology exactly, and exposes common modeling patterns and component combinations in the \texttt{Jutul} ecosystem. Keyword-based search over the same indexed sources supports targeted lookup of API usage and documentation details, and is also useful for discovery of relevant files, examples, and implementation patterns. Together, these retrieval pathways condition the language model prior to code generation and improve alignment with the current simulator API and software version.\looseness=-1

Generated code is first subjected to static analysis before execution. Execution occurs in a local Julia environment, and all runtime errors and solver diagnostics are captured. These diagnostics include constructor mismatches, convergence failures, timestep reductions, and other solver-level signals relevant to physical and numerical admissibility. The agent interprets this feedback to determine whether revisions are required at the level of implementation, parameterization, or missing specification.

In contrast to systems that supplement execution success with heuristic monitoring of physical indicators at runtime \cite{Park2026}, JutulGPT relies on the simulator's own internal enforcement: in JutulDarcy, a simulation does not run to completion unless local and global mass-conservation errors have been reduced below prescribed tolerances and the nonlinear solver has converged, which typically is ensured through the simulators internal timestep control. Simulator completion is therefore itself a certificate of physical validity in the sense of the governing equations, not merely a signal that the code executed without crashing. For reservoir simulation \cite{Lie2019}, interventions such as mesh adaptation or timestep heuristics are strongly problem-dependent and are therefore not abstracted into a generic agent-level policy. The simulator remains the authoritative source of validation, and the agent responds to structured diagnostics rather than duplicating solver-internal logic.\looseness=-1


\section{Case Studies: Evaluating Agentic Model Construction}

The case studies probe the capabilities and limitations of \texttt{JutulGPT} as an agentic interface to scientific simulation. Rather than serving as tutorials or classical benchmarks, they examine how the agent interprets underspecified intent, resolves ambiguity, and converges toward executable, physically consistent models under varying degrees of autonomy. Each case isolates a distinct aspect of agentic model construction, from documentation retrieval to autonomous configuration of heterogeneous multiphysics simulations. While all examples are drawn from porous-media flow, the mechanisms under examination are characteristic of PDE-based simulation more broadly: semantic navigation of documentation, ambiguity detection and resolution, execution-grounded validation, and iterative refinement.

All experiments reported in this section were conducted using Julia~1.12.4, Jutul~0.4.16, and JutulDarcy~0.3.3. In all runs, \texttt{JutulGPT} used OpenAI \texttt{GPT-5.2} with reasoning enabled and greedy decoding (\texttt{temperature=0.0}). Retrieval was performed over searchable indexes built from the locally installed \texttt{JutulDarcy} documentation, docstrings, and example scripts for the active simulator version. Complete agent logs are provided as supplementary material to ensure full transparency and reproducibility. These include tool calls, retrieved context, generated code, execution traces, and repair iterations.

\subsection{Example: Retrieving and exploring documentation}

We first probe the agent's ability to navigate and synthesize simulator documentation. In this mode, the agent does not generate executable models, but retrieves and organizes information from documentation, source files, and example scripts into a coherent summary.

To evaluate this behavior, we issue the following query:
\begin{humanbox}
    What kind of well models can I use in the simulator?
\end{humanbox}
In response, the agent performs a keyword-based search across documentation and source code, retrieves relevant function signatures, and inspects example scripts involving wells:
\begin{toolbox}
    Made a keyword based search for \texttt{well}. Retrieved documentation for 6 functions. Read example ``Introduction to wells''.
\end{toolbox}
The agent organizes the retrieved material into a taxonomy of available well models (Figure~\ref{fig:well-taxonomy}), including vertical wells, multi-segment wells, friction and pressure-drop formulations, and customizable segment-level physics. In addition to listing available constructors, the response explains relationships among model types and indicates typical usage scenarios.

\begin{figure}[tbp]
    \centering
     \begin{agentbox}
        \textbf{Strategy \& Findings}
      
        I searched the API and documentation for well models in JutulDarcy. Here’s a summary of the well modeling options available in the simulator:
        \\
        \textbf{1. Simple Well Model}
        \begin{itemize}
            \item Use \texttt{setup\_well} for a simple well (single cell or trajectory of cells).
            \item Supports basic well parameters: skin, radius, well index (WI), direction, etc.
            \item Suitable for standard well modeling without detailed wellbore flow.
        \end{itemize}
        
        \textbf{2. Multi-Segment Well Model}
        \begin{itemize}
            \item Use \texttt{MultiSegmentWell} or \texttt{setup\_vertical\_well} for a more detailed well representation.
            \item Models the well as a series of connected segments and nodes (multi-segment).
            \item Allows specification of well topology, perforations, and segment models.
            \item Can include a surface node to represent the wellhead.
        \end{itemize}
        
        \textbf{3. Wellbore Flow and Friction Models}
        \begin{itemize}
            \item \texttt{SegmentWellBoreFrictionHB}: Hagedorn and Brown friction model for wellbore segments.
            \item \texttt{MixedWellSegmentFlow}: Two-point approximation for well fluxes.
            \item \texttt{PotentialDropBalanceWell}: Models pressure drop across well segments, including hydrostatic and frictional effects.
        \end{itemize}
        
        \textbf{4. Customization}
        \begin{itemize}
            \item You can specify custom segment models for each segment in a multi-segment well.
            \item The \texttt{setup\_well} and \texttt{setup\_vertical\_well} functions allow for extensive keyword customization (skin, Kh, WI, radius, direction, roughness, etc.).
        \end{itemize}
    \end{agentbox}
    \caption{Taxonomy of well models produced by \texttt{JutulGPT} in response to a documentation query.}
    \label{fig:well-taxonomy}
\end{figure}

This example demonstrates that structured retrieval combined with example inspection enables the agent to synthesize distributed documentation into a unified representation. In particular, the agent surfaces default parameterizations and structural relationships that are not always immediately apparent from isolated docstrings.

At the same time, the example exposes limitations. Many \texttt{Jutul} functions rely on multiple dispatch, where behavior depends on combinations of argument types that are not always explicit in documentation. In such cases, the agent must infer intent from method tables, default keyword arguments, and usage patterns in example scripts. When documentation is sparse or inconsistent, this inference process becomes fragile. The experiment therefore illustrates that documentation structure and example quality directly influence the reliability of agent-mediated exploration.


\subsection{Example: Interpreting and executing a canonical two-phase test case}

We next consider a deliberately simple but physically meaningful test case to establish a baseline for execution-grounded model construction. The quarter five-spot configuration for immiscible two-phase flow in porous media is a canonical problem, with well-known governing equations and qualitative behavior. While the single-phase incompressible version resembles a standard introductory problem in many PDE contexts, the two-phase formulation requires explicit modeling choices, correct use of simulator-specific APIs, and internal consistency across geometry, physical parameters, and numerical controls. This example therefore isolates the core mechanism in a setting with limited structural ambiguity.\looseness=-1

We issue the following modeling request, which specifies the physical scenario and qualitative objective without prescribing implementation details:
\begin{humanbox}
    I want to create two quarter five-spot simulations to illustrate the difference between favourable and unfavourable displacements. Use two-phase immiscible fluids with quadratic relative permeabilities, and inject one pore volume. Please ask for clarifications if you are missing any information. Do not plot the result but run both simulations for the total time.
\end{humanbox}

The agent parses the request into modeling components: (i) quarter five-spot geometry, (ii) immiscible two-phase flow, (iii) quadratic relative permeabilities, and (iv) contrasting mobility ratios corresponding to favourable and unfavourable displacement. The difference between the two regimes is implemented through viscosity assignment, while other components are held fixed to ensure comparability.
\begin{analysisbox}
    \textbf{Creating Julia Simulations}\\
    I need to write Julia code using JutulDarcy and am ready to clarify anything that’s missing. (\dots) There’s the need to compare favorable versus unfavorable displacement, which usually means looking at the mobility ratio based on the viscosities. So, I’m pondering how to set the viscosity ratio correctly for these cases.\smallskip

    \textbf{Setting Up a Five-Spot Grid}\\
    I need to create a quarter five-spot grid, which means I’ll design a 2D square where the injector is located at one corner and the producer at the opposite corner, incorporating symmetry boundaries. I plan to set up a Cartesian grid with Nx and Ny dimensions and implement no-flow boundaries. I'll need to add wells, likely using JutulDarcy. My next steps include consulting examples and utilizing retrieval tools for guidance. (\dots)
\end{analysisbox}

To align implementation with the simulator’s API, the agent invokes retrieval tools to locate relevant examples and documentation:
\begin{toolbox}
    retrieve\_examples(query = "quarter five-spot two-phase immiscible JutulDarcy well corner producer injector")
\end{toolbox}
Using the retrieved material, the agent refines modeling decisions related to geometry, fluid properties, and relative permeability models.
\begin{analysisbox}
    \textbf{Simulating fluid displacements}\\
    I'm creating two simulations (\dots) For favorable displacement, the viscosity ratio should show less viscous water displacing more viscous oil, making $M < 1$. Conversely, if water is more viscous, it creates an unfavorable situation with $M > 1$. (\dots)\smallskip

    \textbf{Setting up quarter five-spot geometry}\\
    I'm working on a quarter five-spot geometry with a square domain. The injector is placed at one corner and the producer at the opposite corner, representing a quarter of a full pattern. I'll use \texttt{CartesianMesh} and set \texttt{nx} and \texttt{ny} to 50. Then, I'll set up the wells with the \texttt{setup\_vertical\_well} function using the correct coordinates. There are also quadratic relative permeabilities to consider, so I need to look up the \texttt{JutulDarcy} relperm API and check the documentation for \texttt{QuadraticRelativePermeability}.
\end{analysisbox}
After several retrieval and refinement iterations, the agent generates an initial implementation:
\begin{agentbox}
\begin{Verbatim}
function quarter_fivespot(; nx=50, L=1000.0, H=10.0, phi=0.3, K=0.1, p0_bar=150.0,
    mu_w=5e-3, mu_o=1e-3,
    rhow=1000.0, rhoo=850.0,
    c_w=1e-10, c_o=1e-10,
    p_bhp_bar=50.0,
    info_level=-1
)
    bar, darcy, kg, meter, day = si_units(:bar, :darcy, :kilogram, :meter, :day)

    mesh = CartesianMesh((nx, nx, 1), (L, L, H))
    domain = reservoir_domain(mesh, permeability = K*darcy, porosity = phi)
    :
\end{Verbatim}
\end{agentbox}
Static analysis using a Julia linter reports no issues:
\begin{toolbox}
    Linter: No issues found.
\end{toolbox}
However, execution produces a runtime error:
\begin{toolbox}
  Code failed!\smallskip

  ERROR: MethodError: no method matching ConstantCompressibilityDensities(::ImmiscibleSystem\{Tuple\{AqueousPhase, LiquidPhase\}, Tuple\{Float64, Float64\}\}; reference\_pressure::Float64, reference\_density::Vector\{Float64\}, compressibility::Vector\{Float64\})\\
  :
\end{toolbox}
Although syntactically valid as Julia code, the constructor invocation does not match the expected API signature. The simulator therefore rejects the configuration. The problematic code
\begin{Verbatim}[fontsize=\footnotesize]
    rho = ConstantCompressibilityDensities(sys, reference_pressure = p0_bar*bar,
                                         reference_density = rhoS, compressibility = [c_w, c_o])
\end{Verbatim}
is revised to
\begin{Verbatim}[fontsize=\footnotesize]
    rho = ConstantCompressibilityDensities(sys, p0_bar*bar, rhoS, [c_w, c_o])
\end{Verbatim}
and the corrected script executes successfully:
\begin{toolbox}
    Code succeeded in 33.2 seconds.
\end{toolbox}

\begin{figure}
    \centering
    \includegraphics[width=0.925\linewidth]{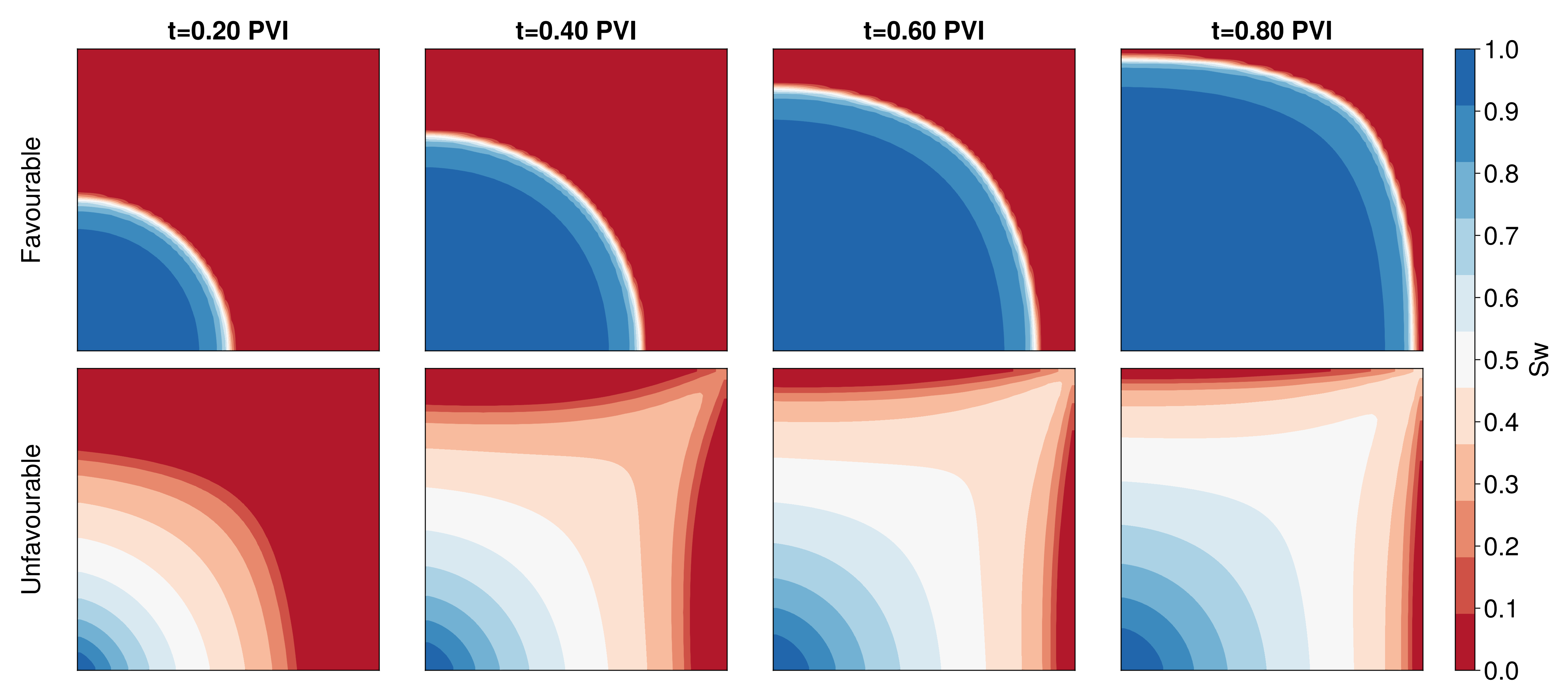}
    \caption{Saturation profiles extracted at fixed fractions of injected pore volume (PVI), providing a normalized basis for comparison independent of injection rate or grid resolution.}
    \label{fig:q5}
\end{figure}

The agent has thus constructed and executed two comparable simulations differing only in mobility ratio. Convergence to a valid configuration occurs through iterative generation, validation, and repair, rather than one-shot code synthesis. Visualization of the resulting saturation fields (Figure~\ref{fig:q5}) is performed using a separate script, as plotting capabilities are not included in the current \texttt{JutulGPT} toolset.

\begin{figure}
    \centering
    \begin{tikzpicture}[
        every node/.style = {draw, gray, rectangle, rounded corners, drop shadow, fill=white}]
    \node[] (p1)
         {\includegraphics[clip,trim=60 150 60 60,width=0.46\textwidth]{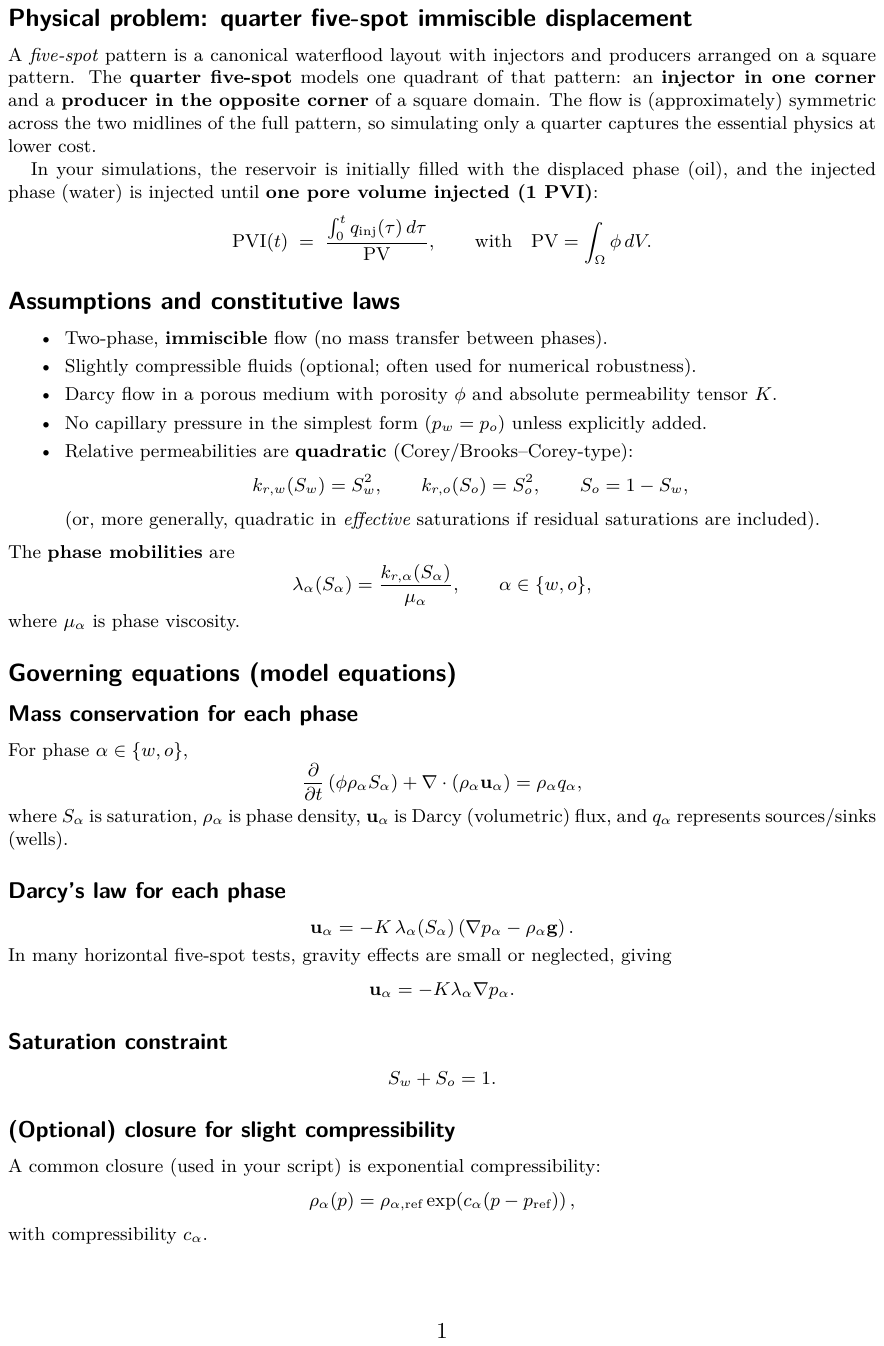}};
    \node[right=1em of p1]
         {\includegraphics[clip,trim=60 150 60 60,width=0.46\textwidth]{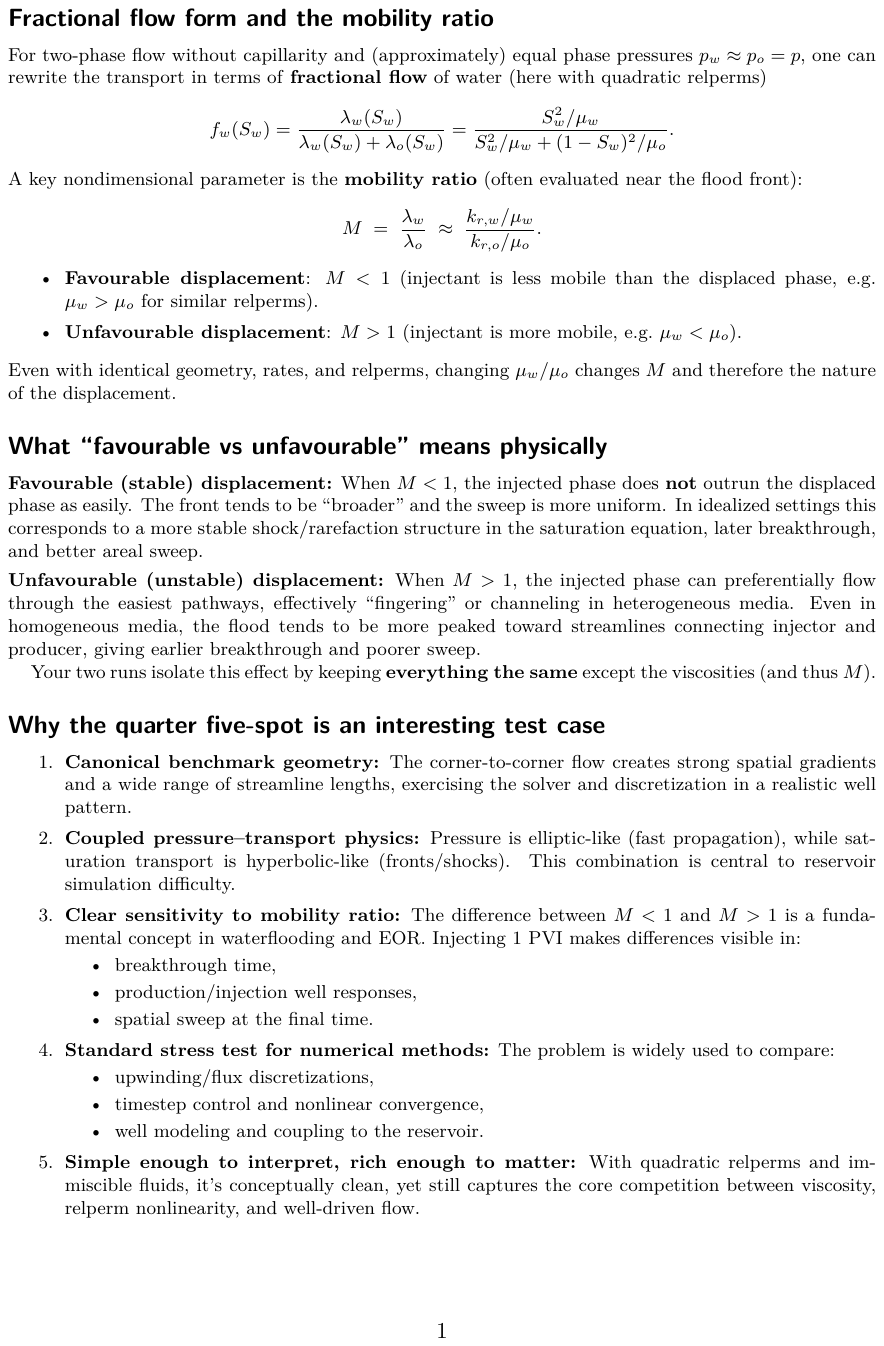}};
    \end{tikzpicture}
    \caption{Explanation generated by the agent when asked to provide a formal description of the simulated problem, including assumptions, governing equations, constitutive laws, introduction of fractional form and mobility ratio, physical explanation of favorable versus unfavorable flow, and brief discussion of the importance of the quarter-five-spot problem.}
    \label{fig:q5-explanation}
\end{figure}

Only after obtaining an executable and validated configuration do we request a formal explanation of the governing equations and physical interpretation (Figure~\ref{fig:q5-explanation}). This ordering emphasizes that explanatory elaboration follows, rather than precedes, executable correctness in the workflow.


\subsection{Example: Interactive construction of a heterogeneous 3D reservoir model}

To examine the agent's capabilities under substantially greater structural complexity, we posed a modeling task defined only at a high conceptual level. The specification includes geometric deformation, stratified heterogeneity, stochastic property assignment, multiple well controls, and a global pore-volume injection constraint, while intentionally leaving several modeling decisions underspecified.

\begin{humanbox}
    Set up a 100x100x30 Cartesian mesh covering a 1000x1000x50 m domain. Immiscible two-phase unfavorable waterflooding with four rate-controlled injectors in the corners and three bhp-controlled producers located inside the domain. Stratigraphy: three equally thick layers with lognormal permeability and arithmetic mean 100,200,900 mD and std 30,60,90 mD. Introduce slight undulating variation in layer thickness. Deform the mesh into an anticline dome shape. Inject in total one pore volume over ten years, equally distributed across all injectors. \smallskip
    
    Consult me for important decisions, and explicitly state all assumptions.
\end{humanbox}

\paragraph{Interpret phase: structured clarification.}
Rather than generating code immediately, the agent identified modeling choices that materially influence physical admissibility or the macroscopic injection constraint. Clarifications were requested for porosity specification, phase viscosities and densities, relative permeability parameters, initial conditions, well configuration details, and deformation amplitudes.

For each underspecified component, the agent proposed explicit defaults together with concise physical justification. Proposed closures included Brooks--Corey relative permeabilities with quadratic exponents and moderate residual saturations, constant phase viscosities yielding an unfavorable mobility ratio, vertical wells perforating all layers, and closed outer boundaries. Stratigraphic undulation and dome deformation were parameterized using smooth sinusoidal and Gaussian functions. Constant porosity was initially proposed, consistent with typical \texttt{JutulDarcy} examples, but later replaced after user override.

Both permeability and porosity were ultimately prescribed as lognormal distributions specified via arithmetic mean and standard deviation. To obtain statistically consistent samples, the agent transformed these descriptors into log-space parameters using moment matching before sampling. This step reconciles user-level statistical inputs with the simulator’s parameterization.

\paragraph{Act phase: geometric deformation and model assembly.}
The structured grid was mapped to undulating stratigraphic interfaces and subsequently deformed into a dome-shaped anticline with zero displacement at the basal boundary. Petrophysical properties were assigned layer-wise using independent lognormal realizations generated with a fixed random seed to ensure reproducibility.

After assembling wells, fluid models, and controls, the total pore volume of the deformed grid was computed explicitly. Injection rates were then derived to ensure that exactly one pore volume would be injected over the ten-year schedule, equally distributed among the four injectors. All wells were implemented as vertical wells perforating the full thickness.

\begin{figure}
    \centering
    \includegraphics[clip,trim=120 0 30 450,width=.65\linewidth]{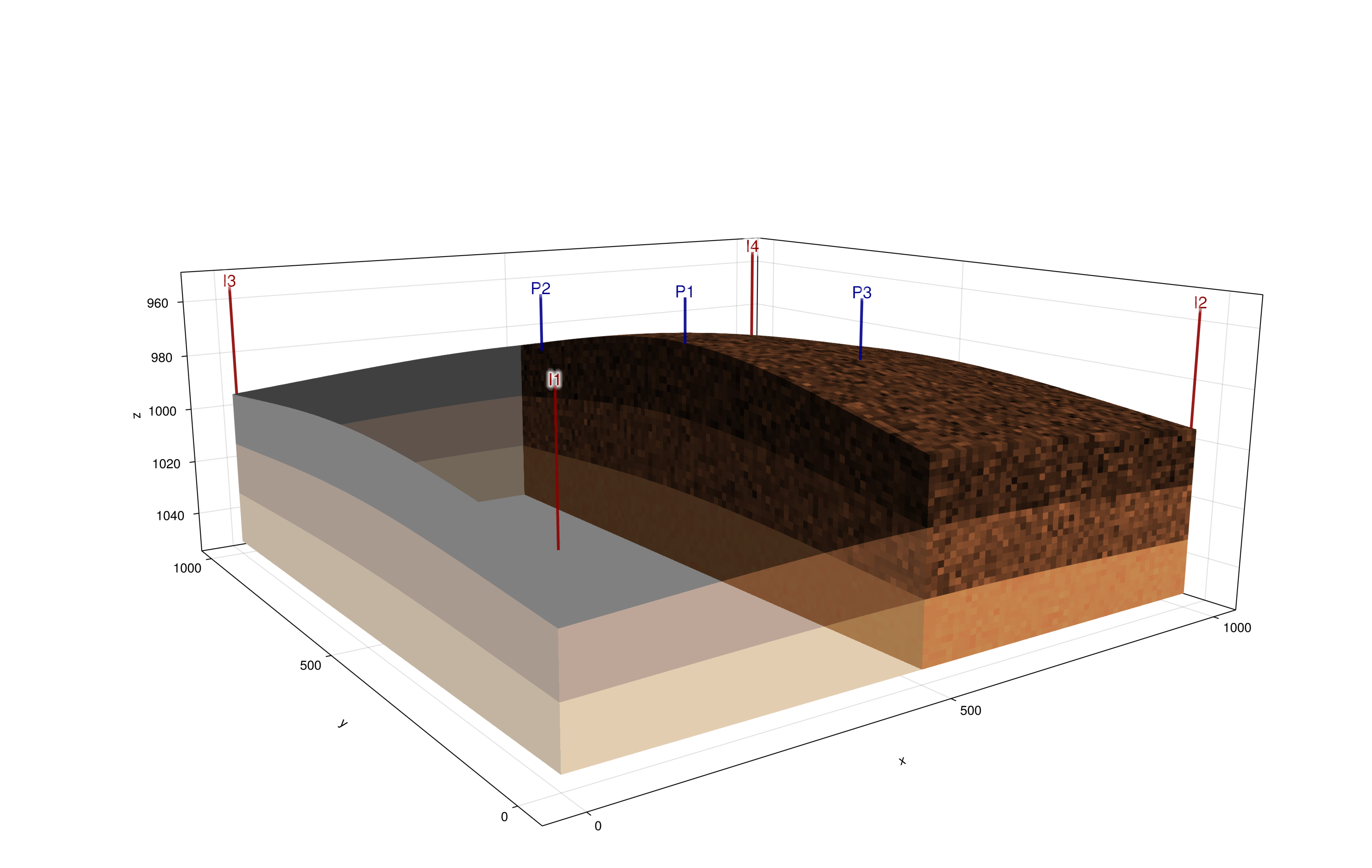}
    \caption{3D reservoir model generated by \texttt{JutulGPT}. The front half uses semi-transparent colors to distinguish the three stratigraphic layers (each with ten grid layers of different average permeability and porosity), while the back half shows log$_10$ of permeability to reveal the stochastic heterogeneity correctly produced within each layer. Four corner injectors (red, I1--I4) and three interior producers (blue, P1--P3) are placed for a peripheral waterflood. In this classic setup, water is injected to displace more viscous oil toward the producers; the anticlinal trap and layered heterogeneity together control sweep efficiency.}
    \label{figure:reference}
\end{figure}

\paragraph{Validate phase: execution-grounded refinement.}
Execution required multiple refinement cycles. Although static analysis reported no syntactic errors, runtime failures arose from constructor mismatches and parameter-placement inconsistencies. These were resolved by inspecting simulator diagnostics and revising the implementation accordingly.

The loop terminated only after successful completion of the full ten-year simulation on the $300{,}000$-cell grid, with verification that the global injection target of one pore volume was satisfied exactly.


\subsection{Example: Autonomous reconstruction from progressively abstract representations} \label{sec:reconstruction}

We use the executable configuration constructed in the previous subsection, together with the associated agent context, as a fully instantiated reference state. In this state, grid topology, geometric deformation, stochastic realizations of petrophysical parameters, fluid closures, well placement and controls, time discretization, and solver settings are all concretely specified.

From this reference configuration, three textual artifacts were generated: (i) a simulator-specific reproduction prompt intended to recreate the model exactly, (ii) a formal technical report expressed in simulator-agnostic terms, and (iii) a compact journal-style description.
\begin{humanbox}
    After the code, using the final setup as reference:
    \begin{enumerate}
        \item Write a formal report-style markdown explanation with LaTeX equations (inline \$...\$, display \$\$...\$\$), using flowing prose (no bullet lists except for equation/boundary condition sequences).
        \item Produce a detailed “reproduction prompt” that would recreate the exact same simulation from scratch.
        \item Provide a shorter journal-style description with key details.
    \end{enumerate}
\end{humanbox}
These artifacts encode the same physical scenario at progressively increasing levels of abstraction. To examine how representational compression affects executable reconstruction, three independent agents were initialized in fully autonomous mode, each provided with one artifact as the sole input. 

In all three cases, the agent produced an executable simulation script that ran successfully and generated physically plausible flow behavior. Each reconstruction yielded a coherent grid, statistically consistent heterogeneous fields, well definitions and control schedules compatible with our initial specification, and stable time integration without runtime failure. Successful execution under simulator validation therefore serves as the minimal admissibility criterion for reconstruction.

Deviations from the reference configuration were nevertheless observed. Because the agents had access only to the textual artifacts and not to the original code or simulation outputs, such deviations reflect admissible interpretations within the degrees of freedom left unspecified by the representation, rather than implementation errors. In standard human workflows, discrepancies of this type would be detected through comparative analysis of geometry, state fields, well responses, and solver diagnostics, followed by iterative refinement. The present framework does not yet automate such multimodal comparison, and systematic reconciliation of reconstructed executable states therefore remains future work.

\paragraph{Reproduction prompt: simulator-specific operational specification.}

The reproduction prompt (Figure~\ref{fig:reproduction-prompt}) was itself generated by the agent in response to the instruction to produce an exact restart specification of the reference configuration. It encodes the model as an operational specification intended to reproduce the simulation exactly, including geometric deformation formulas, statistical parameters, well definitions, time discretization, and a fixed random seed. The prompt explicitly enumerates most numerical degrees of freedom in the executable model, although certain simulator-level closures remain implicit.\looseness=-1

\begin{figure}[ht]
    \centering
    \begin{tikzpicture}[
        every node/.style = {draw,rectangle,rounded corners,drop shadow,inner sep=10pt,fill=white,font=\footnotesize}]
    \node[] (p1)
         {\begin{minipage}{.92\linewidth}
             You are to reproduce *exactly* the same simulation, including all random realizations, geometry deformation, wells, controls, and schedule.
             \begin{enumerate}
                 \item Use Julia with packages `Jutul` and `JutulDarcy` available.
                 \item Set a fixed RNG seed `seed = 12345` using `MersenneTwister(seed)`.
                 \item  Create a Cartesian mesh with dimensions `(100, 100, 30)` covering `(Lx, Ly, Lz) = (1000.0, 1000.0, 50.0)` meters, with origin `(0.0, 0.0, z\_top0)` where `z\_top0 = 1000.0` meters (depth positive downward). Convert it to an unstructured mesh using `UnstructuredMesh`.
                 \item  Deform the mesh nodes:
                 \begin{itemize}
                     \item[--] Define stratigraphic reference interfaces in relative depth as `zmean = (0, Lz/3, 2Lz/3, Lz)`.
                     \item[--] Define undulation with amplitude `A\_und = 2.0 m` and wavelength 
                        `$\lambda$\_und = 500.0 m`: `$\delta$(x,y)=
                        A\_und*sin(2$\pi$*x/$\lambda$\_und)*sin(2$\pi$*y/$\lambda$\_und)`.
                     \item[--] Define two internal interfaces as `z1=zmean[2]+$\delta$`, `z2=zmean[3]-$\delta$` and map 
                        each node’s original relative depth piecewise-linearly from the reference intervals to the undulated intervals.
                     \item[--] Apply an anticline dome with amplitude `A\_dome = 30.0 m` and Gaussian radius 
                        `R\_dome = 400.0 m`: `D(x,y)=A\_dome*exp(-((x-Lx/2)\textasciicircum2+(y-Ly/2)\textasciicircum2)/(2R\_dome\textasciicircum2))`.
                         Shift node depth by `shift=D(x,y)*(1 - z\_rel/Lz)` so the bottom is unchanged
                 \end{itemize}
             \end{enumerate}
         \end{minipage}};
    \end{tikzpicture}
    \caption{Excerpts from the reproduction prompt for the 3D waterflooding case generated by the agent.}
    \label{fig:reproduction-prompt}
\end{figure}

Under this representation, reconstruction fidelity was correspondingly high. Grid geometry, well placement, injection schedule, and deformation parameters were reproduced without structural deviation. Differences arose primarily from the procedure used to sample and assign heterogeneous properties.

Although an identical random seed was specified, deterministic reproduction of a particular spatial realization requires that sampling order and assignment order coincide exactly. In the reference implementation, permeability and porosity were sampled in layer-wise batches, whereas the reconstructed model sampled them cell-by-cell with interleaved draws. The following abridged excerpts illustrate the difference in realization strategy; variable names and surrounding context have been simplified for clarity.
\begin{center}
    \begin{minipage}[T]{.5\linewidth}
       \begin{Verbatim}[fontsize=\footnotesize,frame=leftline]
for l in 1:3
    lognormal!(rng, tmpK, meanK[l], stdK[l])
    lognormal!(rng, tmpPhi, meanPhi[l], stdPhi[l])
    @inbounds for (ix, c) in enumerate(layer_cells)
        K[c]    = tmpK[ix]
        phi[c]  = tmpPhi[ix]
    end
end
       \end{Verbatim}
    \end{minipage}\hspace*{2em}
    \begin{minipage}[T]{.4\linewidth}
       \begin{Verbatim}[fontsize=\footnotesize,frame=leftline]
@inbounds for c in 1:nc
    assign_by_layer(c, mK, sK, mPhi, sPhi)

    Ks = lognormal_sample(rng, mK, sK)
    phis = lognormal_sample(rng, mPhi, sPhi)
    K[c] = (1e-3*Ks)*Darcy
    phi[c] = phis
end
       \end{Verbatim}
    \end{minipage}
\end{center}

Both procedures are statistically consistent with the specified distributions. However, because random numbers are consumed in a different order, the resulting realizations differ under the same seed. In this case, the permeability and porosity fields are i.i.d.\ with no spatial correlation, so the large-scale pressure field and displacement dynamics are governed primarily by geometry, boundary conditions, mobility ratio, and the global injection constraint rather than by coherent preferential pathways. Differences between realizations therefore tend to average out at the scale of the advancing front, even though the local flux partitioning at wells may change. For spatially correlated fields, altered realizations could generate connected high- or low-permeability structures, and reconstruction differences would be expected to have a more pronounced impact on interwell connectivity and the evolution of fluid fronts.

Despite these realization-level differences, the reproduction prompt remains close to an operational specification of the executable state. The remaining degrees of freedom are confined primarily to simulator-level defaults and implementation conventions. In programmatic simulation frameworks such as \texttt{JutulDarcy}, executable state may inherit implicit behavior from library-level closures, multiple dispatch, and constructor defaults. Unless these conventions are explicitly specified, they remain part of the executable configuration without being part of the textual representation.

Comparable effects occur in keyword-driven simulators, where defaulted keywords and ordering conventions introduce implicit state. However, in programmatic interfaces, these dependencies become especially visible when reconstruction is performed via code generation rather than direct reuse of input files.\looseness=-1

\paragraph{Technical report: formal model specification in simulator-agnostic form.}

The formal report, similar in structure to Figure~\ref{fig:q5-explanation}, preserved governing equations, geometric definitions, statistical parameterizations, grid dimensions, well controls, and global injection constraints, but did not specify the density--pressure relationship.

\begin{figure}
    \centering
    \includegraphics[clip,trim=120 0 30 450,width=.475\textwidth]{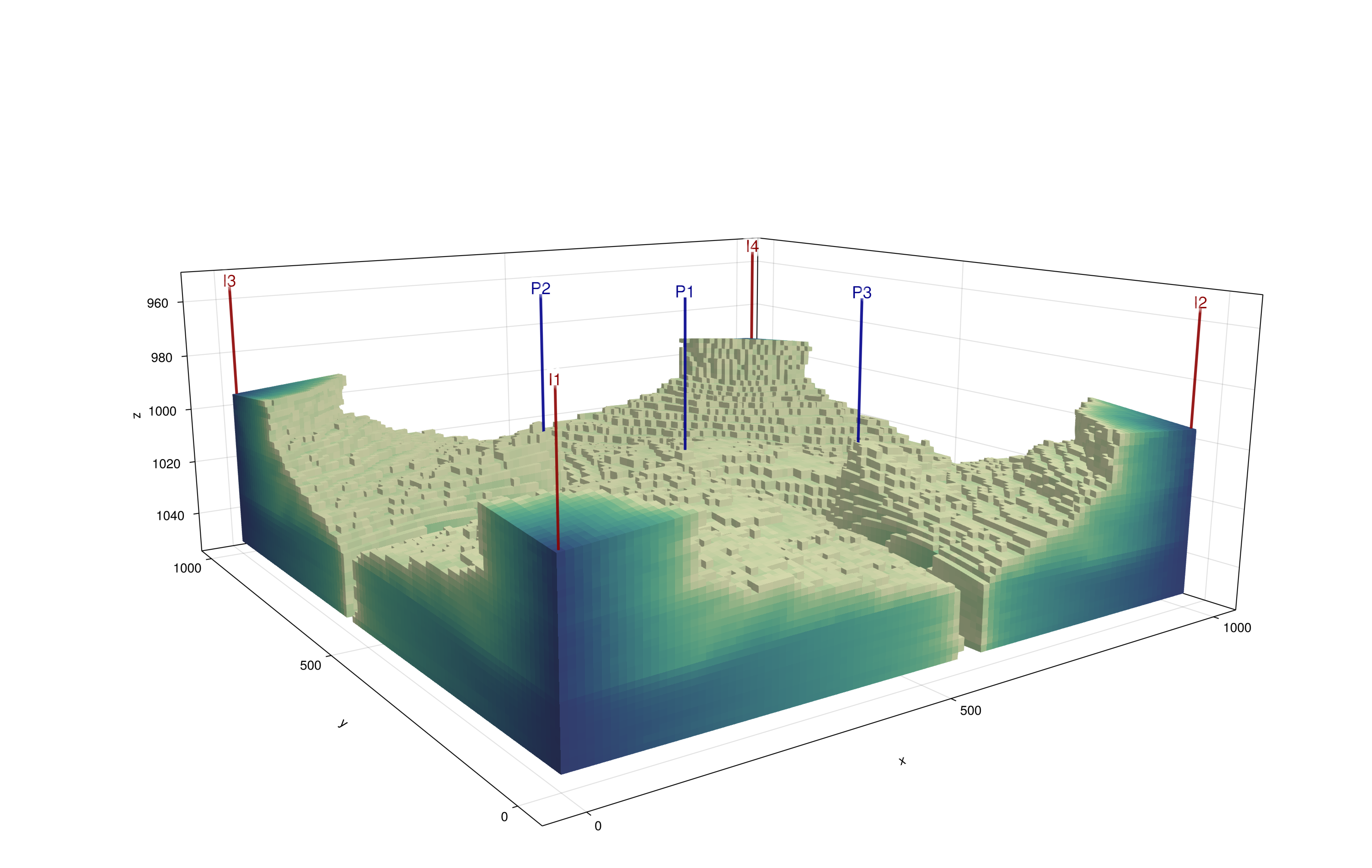}
    \includegraphics[clip,trim=120 0 30 450,width=.475\textwidth]{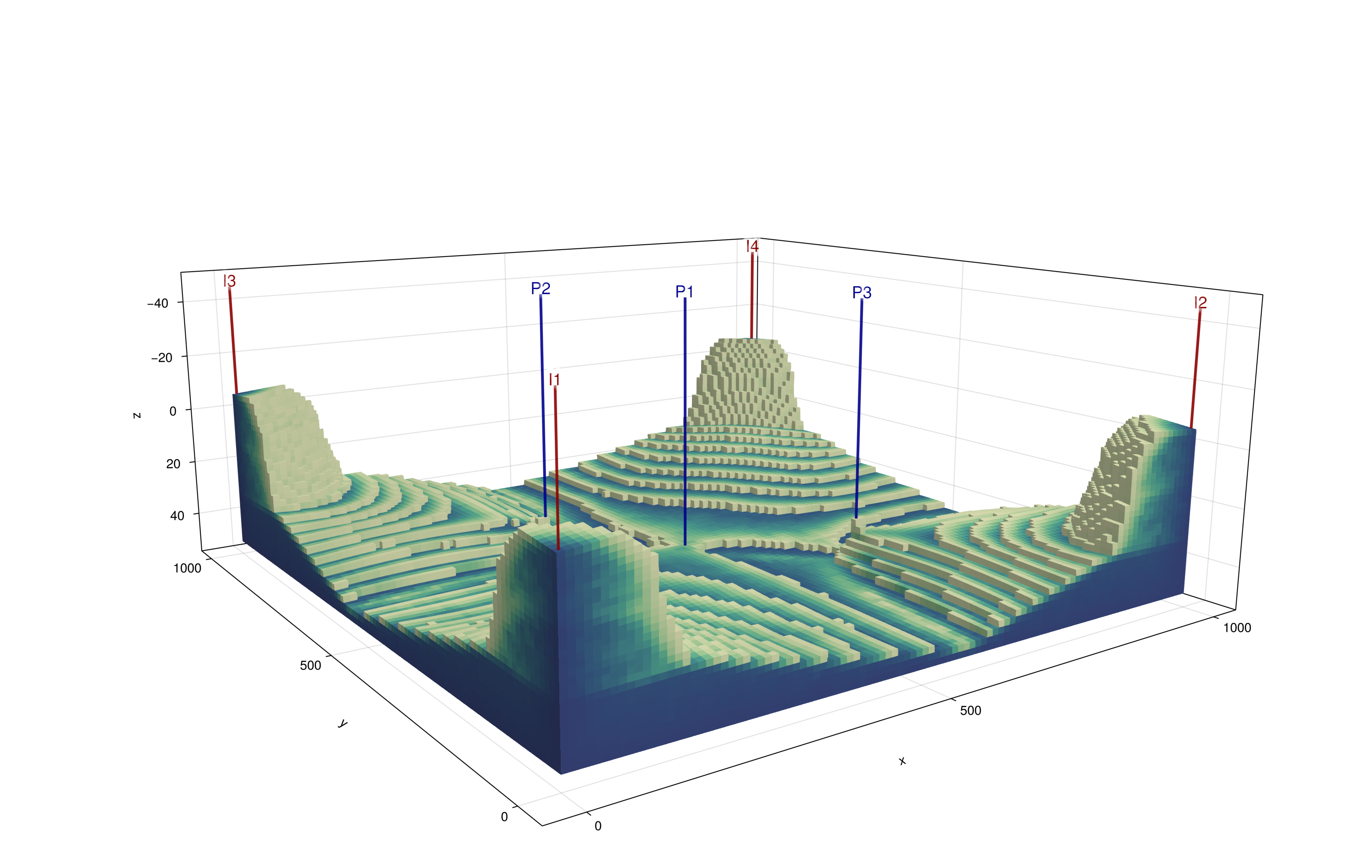}
    \caption{Water saturation for the compressible reference case (left) and the incompressible reproduction from the technical report (right). Densities are specified at surface conditions (1\,bar), but the simulation runs at reservoir pressures (88--90\,bar). In the reference case, higher oil compressibility substantially reduces the effective density contrast and weakens buoyancy; the diffusive pressure response slows vertical redistribution, allowing viscous fingering to produce a mixed, voxelated pattern. In contrast, the incompressible case maintains the full surface density contrast, enabling rapid buoyancy segregation: denser water quickly underrides lighter oil and channels along the high-permeability bottom layer, forming sharp, bottom-conformant fronts.\looseness=-1}
    \label{fig:report-differences}
\end{figure}
The reason it was absent is instructive. In the reference implementation, compressibility was never a decision the agent made explicitly: it wrote code that invoked a simulator default, and that default was outside the assumption space the agent was reasoning over. A choice that was never made cannot appear in an assumption log, and a choice that was never logged cannot appear in any representation derived from that log. The reconstruction agent therefore encountered a genuinely underspecified closure and resolved it explicitly by setting zero compressibility, whereas the reference had silently inherited a compressible model. This resulted in significantly different fluid behavior, as shown in Figure~\ref{fig:report-differences}.

The divergence thus illustrates two compounding effects: latent under-specification in the representation, and a structural gap in the current architecture whereby tacit assumptions---choices inherited from simulator defaults, rather than made explicitly by the agent---are invisible to the assumption log and therefore to any downstream representation.

Fortunately, this divergence can be corrected by modifying a localized constitutive definition in the reconstruction script. In the report-based reconstruction, the incompressible formulation was introduced explicitly through the following readily identifiable statements:
\begin{Verbatim}[fontsize=\footnotesize]
   # Constant densities (no compressibility)
   rho = ConstantCompressibilityDensities(sys, 1.0*bar, rhoS, [0.0, 0.0])
   replace_variables!(model, PhaseMassDensities = rho)
\end{Verbatim}
Restoring the reference behavior therefore requires only removal (or modification) of this closure assignment. The discrepancy is thus confined to a single constitutive component and does not affect geometry, discretization, or structural model assembly.

It is worth noting that this particular tacit assumption is unusually easy to detect: the reconstruction agent happened to make its choice explicitly, leaving a visible two-line signature in the code. The reference agent's compressible default, by contrast, leaves no corresponding signature; it is simply absent from the code. In general, tacit assumptions originating from simulator defaults are harder to audit precisely because they produce no code to inspect.

As for the reproduction prompt, permeability and porosity were sampled cell-by-cell with interleaved random draws. This differs from the layer-batched sampling strategy used in the reference implementation and produces a distinct spatial realization despite identical statistical descriptors and seed. In addition, the reconstructed model explicitly selected a CSR linear-algebra backend and a BiCGStab solver with CPR-type preconditioning, whereas the reference relied on simulator defaults. These choices are numerically consistent with the problem formulation but introduce implementation-level differences that may affect convergence behavior and, consequently, transient solution details.

\paragraph{Journal-style description: compact conceptual specification.}

The journal-style description (Figure~\ref{fig:journal-prompt}) encodes the physical scenario at a higher level of abstraction, summarizing geometry, stratigraphy, heterogeneity statistics, fluid properties, and injection strategy without specifying all detail. It preserves the conceptual intent of a three-layer stratigraphy, lognormal heterogeneity, anticline deformation, and injection of one pore volume over ten years, but omits several operational and procedural parameters.\looseness=-1

\begin{figure}[ht]
    \centering
    \begin{tikzpicture}[
        every node/.style = {draw,rectangle,rounded corners,drop shadow,inner sep=10pt,fill=white,font=\footnotesize}]
    \node[] (p1)
         {\begin{minipage}{.9\linewidth}
             We simulate an immiscible, two-phase unfavorable waterflood in a deformed 3D $1000\times 1000\times 50\ \mathrm{m}$ reservoir discretized by a $100\times 100\times 30$ grid (300k cells). Stratigraphy comprises three units (10 layers each) with i.i.d. lognormal permeability fields having arithmetic means of 100, 200, and 900 mD and standard deviations of 30, 60, and 90 mD, respectively; porosity is i.i.d. lognormal with means 0.18/0.20/0.22 and standard deviations equal to one quarter of the mean. Layer interfaces are laterally undulated (2 m amplitude) and the grid is further deformed into a dome-shaped anticline (30 m top uplift). Water–oil flow uses constant densities (1000/800 kg$\cdot$m$^{-3}$), constant viscosities (0.5/5 cP), and Brooks–Corey relative permeability with exponents 2 and residual saturations 0.2/0.2. Four corner injectors are rate-controlled to inject one pore volume over 10 years (equal split), while three interior producers are BHP-controlled at 50 bar; the initial state is uniform at 150 bar and $S_w=0.2$.
         \end{minipage}};
    \end{tikzpicture}
    \caption{Journal-style description of the 3D waterflooding case generated by the agent.}
    \label{fig:journal-prompt}
\end{figure}

In contrast to the report-based reconstruction, where divergence primarily arose from a localized constitutive choice, the journal-based reconstruction exhibits three first-order differences relative to the reference configuration: (i) adoption of an incompressible density closure, (ii) reinterpretation of interior well placement, and (iii) altered geometric deformation, which changes the total bulk and pore volume of the model.

First, the reconstruction again selected a constant-density formulation. As discussed for the report case, this choice is admissible under the representation but modifies the governing model and directly affects pressure evolution and rate decay behavior. However, as shown above, it can be corrected through a localized closure modification.

Second, the interior producer locations were reinterpreted into a symmetric configuration that differs from the original coordinates. Under BHP control, well placement directly influences drainage geometry and rate partitioning among producers. Even with statistically identical i.i.d.\ permeability fields, such changes can produce a substantially different distribution of production rates.

\begin{figure}
    \centering
    \includegraphics[width=.95\linewidth]{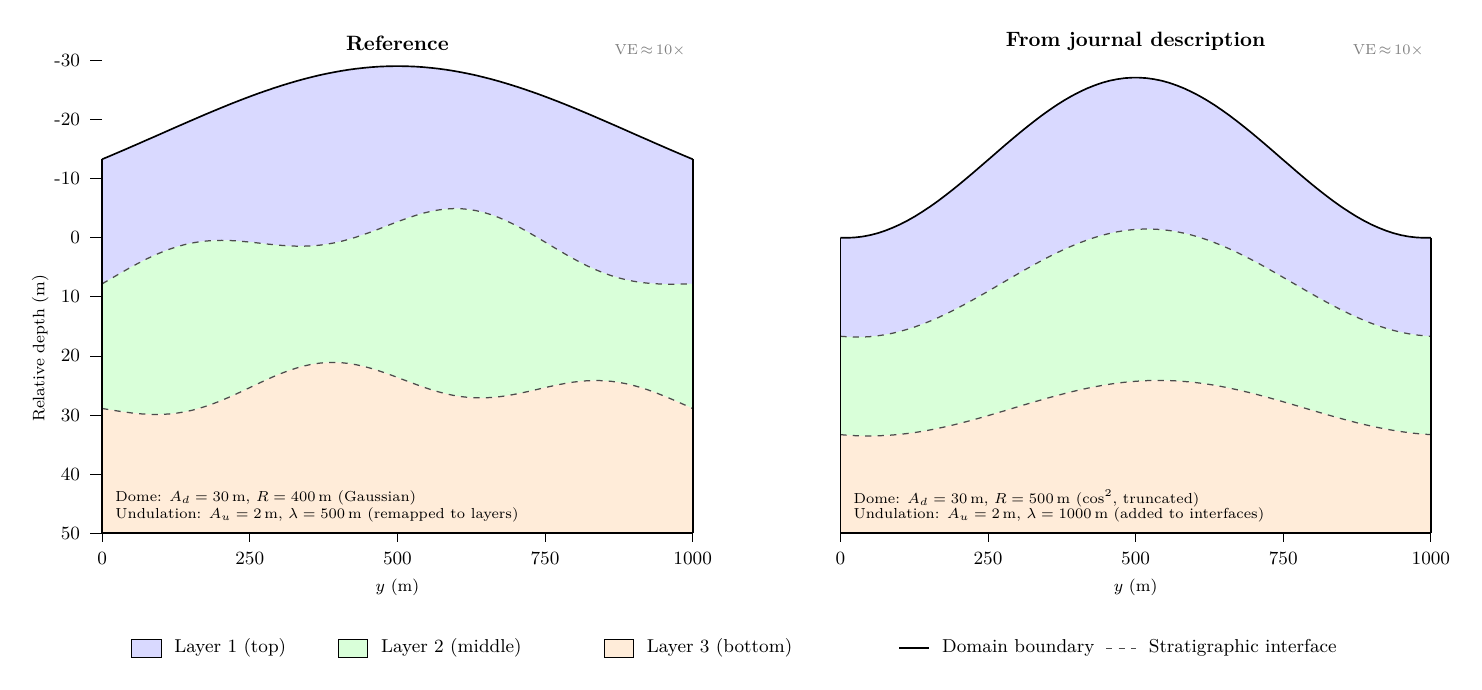}
    \caption{Comparison of vertical cross-sections through $x=400$ for the reference model and the model reconstructed from the journal-style description.}
    \label{fig:journal-differences}
\end{figure}

Third, the dome deformation and stratigraphic undulation were reconstructed using different functional forms (Figure~\ref{fig:journal-differences}). While property assignment and pore-volume computation algorithms remain unchanged, the altered geometry changes the total pore volume of the reservoir. This directly changes both the amount of fluid initially contained in the system and the total amount injected over the simulation period to satisfy the one-pore-volume requirement.

At the implementation level, sampling order and solver configuration again differed, but these represent secondary procedural variations relative to the first-order effects described above.

At this level of abstraction, reconstruction variability shifts from constitutive detail to architectural interpretation, producing executable models that remain conceptually consistent yet structurally distinct from the reference configuration.

\paragraph{Synthesis: representational compression and executable state.}

Across all three reconstruction experiments, the agent produced simulation scripts that executed successfully and generated physically plausible flow behavior. The deviations relative to the reference configuration therefore reflect variability among admissible reconstructions rather than failures of model construction.

The experiments demonstrate that an executable simulator configuration contains more information than is typically conveyed by textual representations. The reproduction prompt approaches an operational specification and therefore tightly constrains reconstruction, although certain simulator-level defaults and procedural conventions remain implicit. The technical report preserves governing equations, parameter values, and geometric definitions but abstracts away implementation context, leaving implicit constitutive closures and procedural details underdetermined. The journal-style description further compresses the representation, omitting structural, configurational, and discretization details and thereby enlarging the admissible reconstruction space.

Importantly, some ambiguities, such as the density closure, were already present in the original specification and could have manifested under repeated reconstructions even from the reproduction prompt. Differences across reconstructions therefore expose latent degrees of freedom in the model rather than errors in interpretation. Reconstruction variability makes implicit modeling assumptions visible.

In all cases, the interpret--act--validate loop grounds reconstruction in executable feedback, but it cannot resolve ambiguities that are not encoded in the representation itself. Deterministic reproducibility therefore requires explicit specification not only of governing equations and parameter values, but also of constitutive closures, stochastic realization procedures, and discretization choices. When these elements (and possible variability) are documented explicitly and illustrated consistently, the admissible reconstruction space narrows and variability is reduced.

The difficulty of reproducing computational results from textual descriptions alone is a long-recognized challenge in scientific computing \cite{LeVeque2007, LeVeque2012}: parameter values, function invocation sequences, and other computational details are typically omitted from published descriptions but are critical for replicating results. A related phenomenon has been observed in large community benchmarks. For example, the 11th Society of Petroleum Engineers Comparative Solution Project (SPE11) for geological CO$_2$ storage \cite{Nordbotten2024} was defined with an unusually detailed problem description and supporting input resources intended to minimize ambiguity. Nevertheless, subsequent intercomparison revealed that variations in results from the 18 contributing teams were strongly influenced by factors not documented in the technical responses, and that unreported human setup choices were at least as impactful as the documented computational choices \cite{Nordbotten2025}. 

The reconstruction experiments above demonstrate the same phenomenon in the context of agent-mediated model construction: even carefully controlled textual specifications do not fully determine executable simulator state, and variability in interpretation and reconstruction remains intrinsic to computational modeling workflows. More broadly, this highlights a general limitation of text-only reporting in simulation studies: without accompanying code and input artifacts, published descriptions seldom determine the computational experiment setup uniquely, making reliable reconstruction difficult.


\section{Discussion and Outlook: Designing Agentic Simulation Environments}
\label{sec:discussion}

The experiments demonstrate that execution-grounded agentic workflows can mediate interaction with high-fidelity simulators while preserving physical and numerical admissibility. At the same time, they expose two structural limitations that constrain reliability: the invisibility of tacit assumptions inherited from simulator defaults, and the irreducible information loss between executable state and its textual representations. The subsections below examine these limitations and their implications for workflow design, reproducibility, and the architecture of future systems.

\subsection{Current limitations}

Although \texttt{JutulGPT} can autonomously construct and execute multiphase models from high-level specifications, its reasoning depth is constrained by context length and prompt budget. As simulation complexity increases, maintaining coherence across geometry, properties, well configuration, and solver setup becomes increasingly demanding.

Reconstruction reliability depends strongly on documentation structure. Functions lacking explicit docstrings, typed signatures, or illustrative examples are effectively opaque to the agent. Conversely, well-structured documentation narrows the admissible reconstruction space. The experiments therefore indicate that documentation quality directly influences agent-mediated reproducibility.

A related limitation concerns tacit assumptions introduced through simulator defaults. The assumption log records choices the agent makes explicitly: selecting a value, resolving an ambiguity, responding to a clarification query. It does not capture choices the agent makes tacitly, by writing code that invokes simulator defaults without those defaults ever being considered as modelling decisions. The compressibility divergence in the reconstruction experiments is an instance of this: the agent wrote a density-model constructor whose default was compressible, and that default was never part of the agent's assumption space. Closing this gap would require the agent to maintain a checklist of choices that a given simulator configuration leaves to defaults, and to either log each default explicitly or prompt the user to confirm it.\looseness=-1

A further limitation concerns output analysis. While the agent can construct and execute models, it does not yet perform structured comparison of simulation outputs. In conventional workflows, discrepancies in physical quantities of interest trigger iterative refinement. Enabling systematic multimodal output comparison, both human-guided and automated, remains an important development direction.\looseness=-1

\subsection{Implications for workflow design}

Agentic interfaces alter the structure of simulation workflows. Rather than manually encoding configurations, users specify intent in natural language, while the agent performs retrieval, translation, and validation. The simulator remains the authoritative computational core, and executable feedback constrains the agent’s reasoning.

This shift requires deliberate redesign of simulator outputs and documentation. Logs and diagnostics must be structured and machine-readable. Documentation must expose semantic relationships rather than isolated function descriptions. Decisions regarding fully autonomous execution versus supervised interaction affect transparency and reproducibility and must be matched to task criticality.

\subsection{Reproducibility and executable state}

The reconstruction experiments underscore that reproducibility of a simulation requires preservation of its executable state: the generated code together with all its necessary parameters and pinned versions of the simulator and its dependencies. Natural-language descriptions alone do not uniquely determine this state, as the reconstruction experiments demonstrate.

Separately, reproducibility of the agent's construction process, i.e., understanding which choices were made, why, and from what retrieved context, requires the broader traceability record: prompts, retrieved documentation fragments, assumption logs, and repair iterations. This record allows divergences arising from explicit modeling choices to be diagnosed and corrected, but it cannot surface choices that were never part of the agent's assumption space.

\texttt{JutulGPT} maintains this traceability record by construction as an audit trail. However, as demonstrated in Section~\ref{sec:reconstruction}, the record cannot compensate for ambiguities in the executable state itself: when constitutive closures or structural assumptions are left unspecified, different agents will make different admissible choices, and the resulting simulations will diverge regardless of how thoroughly the construction process is logged.

\subsection{Simulator--agent interaction}

Effective deployment of agentic interfaces requires structured communication between the simulator, agent, and user. The experiments demonstrate that LLMs can interpret unstructured simulator diagnostics directly, one of their genuine strengths, but this parsing is inherently fragile when error messages are terse, version-dependent, or domain-specific. Structured, machine-readable diagnostic formats would make the validate phase more robust and less dependent on the LLM's natural-language parsing capability, and represent a worthwhile design investment for simulators targeting agent-assisted workflows.

The appropriate level of autonomy depends on context. Exploratory modeling may tolerate greater independence, whereas safety-critical simulations demand human oversight. The simulator remains the authoritative source of physical truth; the agent functions as translator and orchestrator; and the human retains conceptual responsibility.


\section{Conclusions}
\label{sec:conclusion}
The results demonstrate that execution-grounded agentic interfaces can mediate interaction with high-fidelity simulators without replacing them. The central principle is that the simulator remains the authoritative computational engine, while the agent manages interpretation, retrieval, and orchestration.

Several directions are natural extensions of the present work. Because \texttt{Jutul} is designed around residual-form equations with automatic differentiation permeating the entire computational stack, sensitivities of simulation outputs with respect to any model parameter (including constitutive closure coefficients, grid spacing, and timestep sizes) are available without hand-coded adjoint interfaces. This means that the same agent that constructs a forward model could formulate and solve inverse problems (history matching, parameter estimation, sensitivity analysis) within a unified execution-grounded loop. More broadly, the modular architecture of \texttt{Jutul} positions the agent as an orchestration layer across coupled physical processes, where differentiability provides access to cross-module sensitivities that would otherwise require bespoke derivation.

A complementary direction is to expose simulator functionality through standardized tool interfaces, allowing general-purpose agents to interact with domain-specific simulation capabilities without requiring a fully custom implementation. In such a setup, documentation retrieval, execution environments, and structured diagnostic parsing could coexist with general reasoning models, lowering adoption barriers while retaining the depth of integration that execution-grounded validation requires.

The central open problem is closing the tacit-assumption gap: ensuring that modeling choices inherited implicitly through constructor defaults and library conventions become as visible and auditable as the choices the agent makes explicitly.

\section*{Acknowledgments}
This research was conducted within SINTEF Digital’s Agent Lab. 

\printbibliography 

\end{document}